\documentclass[10pt,twocolumn,a4paper]{IEEEtran}
\usepackage{graphicx}
\usepackage{epsfig,amsfonts,amsmath,amssymb,amstext,amsthm}
\usepackage{threeparttable}
\usepackage{epstopdf}
\usepackage{fancyhdr}
\usepackage{footnote}
\usepackage{setspace}
\usepackage{wrapfig,balance,url}
\newcommand{\bm}[1]{\mbox{\pmb{$#1$}}}

\begin{document}
\title{\Huge{Doppler Effect Assisted Wireless Communication for Interference Mitigation}}
\author{Dushyantha A. Basnayaka,~\IEEEmembership{Senior Member,~IEEE}, and Tharmalingam Ratnarajah,~\IEEEmembership{Senior Member,~IEEE}
\thanks{Authors are with Institute for Digital Communication, University of Edinburgh, United Kingdom (e-mail: \{d.basnayaka\}@ed.ac.uk).}
}
\maketitle
\begin{abstract}
Doppler effect is a fundamental phenomenon that appears in wave propagation, where a moving observer experiences dilation or contraction of wavelength of a wave. It also appears in radio frequency (RF) wireless communication when there exists a relative movement between the transmitter and the receiver, and is widely considered as a major impairment for reliable wireless communication. The current paper proposes Doppler Assisted Wireless Communication (DAWC), that exploits Doppler effect and uses kinetic energy for co-channel interference (CCI) mitigation. The proposed system also exploits the propagation environment and the network topology, and consists of an access point (AP) with a rotating drum antenna. The rotating drum receive antenna is designed in such a way that it shifts the interfering signals away from the desired signal band. This paper includes a detailed system model, and the results show that under favorable fading conditions, co-channel interference can be significantly reduced. Therefore, it is anticipated that more sophisticated wireless systems and networks can be designed by extending the basic ideas proposed herein.    
\end{abstract}
\vspace{-2mm}
\section{Introduction}\label{sec:introduction}
\begin{figure*}[t]
	\centering
	\includegraphics*[scale=2.1]{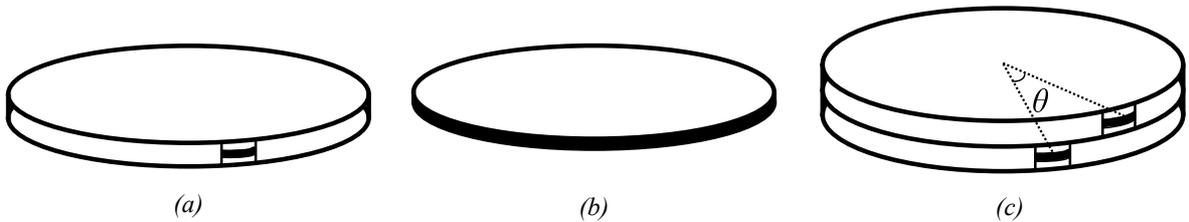}
	\caption{A possible antenna implementation for Doppler Assisted Interference Mitigation (DAIM), where (a) a fixed antenna canister with an opening, where a small portion of the antenna is visible to outside, (b) a rotating thin drum antenna, where the receiving surface is located on the outer curve surface, (c) a possible implementation to receive two desired signals from two transmitters of azimuth separation of $\theta$. Note that the rotating drum antenna is located inside the fixed canister.}
	\label{fig:con_en:fig3}
\end{figure*}

A multitude of wireless networks have revolutionized the modern living for a half a century, and are expected to revolutionize our lives in an unprecedented scale in the future too \cite{Osse14}. Todays' wireless networks--often digital wireless networks--are used for various activities such as mass communication, security and surveillance systems, sensing and disaster monitoring networks, satellite communication and tactical communication systems. Wireless networks typically consist of a collection of wireless transmitters and receivers, and use a fixed band of radio frequency (RF) spectrum for communication. Due to spectrum scarcity, often wireless networks reuse their limited frequency spectrum, which in turn gives rise to a fundamental problem in wireless communication known as co-channel interference \cite{Zander92}.\\
\indent Co-channel interference (CCI) occurs when wireless stations that are nearby use/reuse overlapping spectrum. Modern wireless networks widely employ many intelligent and adaptive physical (PHY) layer interference mitigation/avoidance/management techniques such as interference detection and subtraction (also known as successive interference cancellation or SIC), coordinated multi-point systems (CoMP), interference alignment (IA), diversity receivers (such as maximal ratio combining (MRC), zero forcing (ZF) and minimum mean-squared error (MMSE)), cognitive radio (CR), cooperative communication and transmit power control. Often PHY layer techniques are more effective, but multiple access control (MAC) layer techniques such as schedule randomization, measurement and rescheduling, and super controller also exist \cite{PaGo09}. Furthermore, cross layer techniques that combine and jointly optimize two or more MAC and PHY techniques for interference mitigation are also widely considered \cite{Ch11}.\\
\indent The PHY techniques add (or are expected to add) significant intelligence to future wireless networks. For instance, non-orthogonal multiple access systems (NOMA), which uses interference detection and subtraction along with power control is expected to increase the spectral efficiency and the system throughput in 5th generation (5G) mass communication networks \cite{Di17, SaKiBe13}. In CoMP, a number of co-channel transmitters provide coordinated transmission to multiple receivers and multiple receivers provide coordinated reception to multiple co-channel transmitters \cite{MaGrFe11}. In IA, all the co-channel transmitters cooperatively align--by exploiting channel state information (CSI) at transmitters--their transmissions in such a way that the interference subspaces at all the receivers jointly are limited to a smaller dimensional subspace, and is orthogonal to the desired signal subspace \cite{ElLoHe12}. Diversity receivers use multi-antenna techniques for interference mitigation \cite{Godara97}, and CR learns from the environment, and adapts its transmission. If a particular channel is occupied by a primary user, CR halts transmission or transmits at a lower power level so that the possible interference to primary user is minimized \cite{Haykin05}.\\        
\indent The focus of this paper is also interference mitigation at PHY layer, where we envision to add a new degree of freedom to existing wireless receivers by exploiting Doppler effect. We introduce a new paradigm for wireless communication, namely Doppler Assisted Wireless Communication (DAWC in short) \cite{Dush_patent}, and consider a receiver or an access point (AP) in a typical wireless sensor network with a circular high speed rotating drum antenna as shown in Fig. \ref{fig:con_en:fig3}. 
\subsection{Rotating Drum Antenna}\label{sec:rotating-drum-antenna}
Fig. \ref{fig:con_en:fig3} illustrates a circular drum antenna considered herein for Doppler assisted wireless communication. In order to exploit Doppler effect, the AP has a circular antenna, which consists of two parts. 1) a fixed circular canister 2) a rotating drum antenna as shown in Fig. \ref{fig:con_en:fig3}-(b). The radiating surface of the antenna lies on the curve surface of the drum, and the rotating drum antenna is located inside the circular cannister. There is an opening on the curve surface of the canister, and it is directed to the direction of the desired transmitter. Consequently, the channel responses of the desired link and the other interfering links which have reasonable azimuth separation exhibit different frequency characteristics. Despite Doppler effect being considered as an major impairment, the results in this paper indicate that successful receiver side CCI mitigation techniques (henceforth DAIM to denote Doppler Assisted Interference Mitigation) can still be developed based on a rotating drum antenna and Doppler effect.\\
\indent The round and/or rotating antenna units are used in several key widely-used systems such as TACAN (for tactical air navigation) and LIDAR (for light detection and ranging). The TACAN system has multiple antennas installed in a circle, and electronically steers a radio beam in order to provide directional information for distance targets over a $360^o$ azimuth. The radio beam rotates (often electronically) merely to serve targets located around it. However, the rotating antenna in Doppler assisted wireless communication is a key unit that gives rise to Doppler effect, and fundamentally important for the operation of the system. It typically rotates at very high speed than the beam in TACAN system. Furthermore, modern autonomous cars also employ a system known as LIDAR, which also has a rotating unit. LIDAR is a variation of conventional RADAR, and uses laser light instead of radiowaves to make high-resolution topological maps around automobiles. LIDAR antenna rotates in order merely to map $360^o$ angle, and not for any other fundamental reason. The TACAN and LIDAR systems are hence fundamentally different from the system proposed in the current paper, and tackle entirely different challenges.\\
\indent MIMO (or its more popular variant, massive MIMO) is the state-of-the-art for CCI mitigation, but heavily relies on CSI \cite{Lu14}. The proposed system does not rely on CSI for CCI mitigation (note however that, it still uses CSI of the desired user for data detection). Typically, CCI may occur from a single co-channel transmitter or numerous co-channel transmitters. If CCI occurs from multiple co-channel users, MIMO systems need fairly accurate CSI of all co-channel users for successful CCI mitigation \cite{Paulraj03}. They let the interference into the system, and uses ever more complex signal processing techniques (a software domain approach) for CCI mitigation. In essence, massive MIMO lets the enemy (metaphorically to denote CCI) into its own backyard, and fight head-on. In contrast, the proposed system can handle any number of interferers, and automatically suppresses co-channel multi-path signals with a reasonable azimuth separation to the desired multi-paths even before they corrupt the desired signal. In that sense, DAWC is a paradigm shifting technology, which keeps the enemy at the bay.\\
\indent Furthermore, note that the results in this paper are applicable to wireless networks based on microwave, mmwave frequencies, and also to coherent optical laser communication systems \cite{KoKo90, Gr10}.\\       
\indent This paper includes a detailed study of Doppler assisted wireless communication based on a detailed digital communication system simulation on MATLAB. The rest of the paper also includes the system model in Sec. \ref{sec:system}, performance analysis and discussions in Sec. \ref{sec:performance}, further remarks in Sec. \ref{sec:further-remarks}, and conclusions in Sec. \ref{sec:conclusions}. 
\begin{table}[t]
	\caption{Notations and their definitions}
	\centering
	\begin{tabular}{| l | c |}
		\hline \hline
		Definition & Symbol \\
		\hline
		Carrier frequency & $f_c$ \\
		Speed of light & C\\
		Sampling frequency & $F_s$\\
		Signal bandwidth & $B_w$\\
		Symbol duration & $T_s$\\
		Rician factor & $K$\\
		Number of $S/I$ multi-path components & $N_S/N_I$\\
		Amplitute of the $n$th $S/I$ multi-path & $\alpha_n^S/\alpha_n^I$\\
		Phase of the $n$th $S/I$ multi-path & $\phi_n^S/\phi_n^I$\\
		Delay of the $n$th $S/I$ multi-path & $\tau_n^S/\tau_n^I$\\
		Azimuthal separation of $S$ and $I$ & $\theta$\\
		\hline
	\end{tabular}\label{table:sm:1}
\end{table}
\section{System Model}\label{sec:system}
\begin{figure*}[t]
	\centerline{\includegraphics*[scale=2.2]{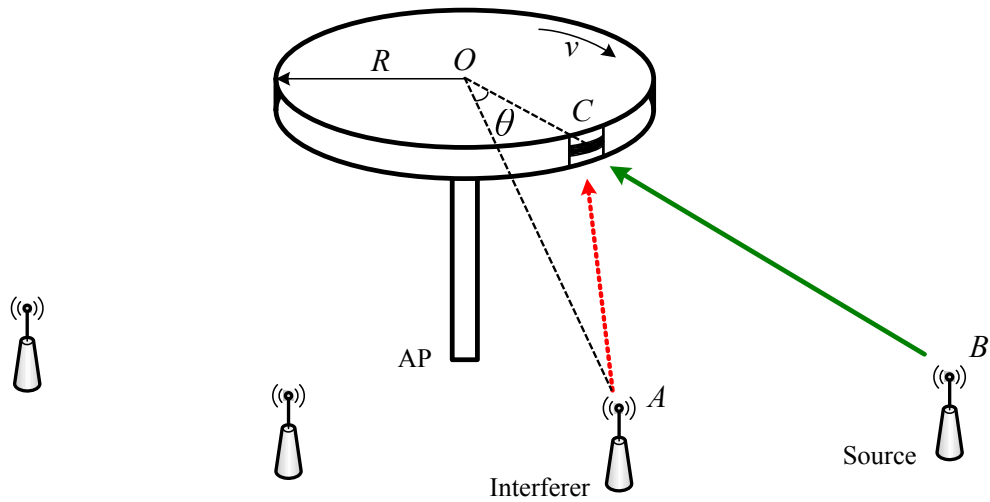}}
	\caption{A possible implementation of a spinning antenna wireless AP for Doppler assisted wireless communication. The figure is not to the scale, and AP is exaggerated for exposition. To reduce the clutter, not all wireless links are shown.}
	\label{fig:con_en:fig2}
\end{figure*}
A narrow-band uplink communication from a wireless station, $S$ (to denote source) to AP in a wireless network is considered, where another wireless station, $I$ (to denote the interferer) located at an azimuthal angle separation of $\theta$ poses co-channel interference as shown in Fig. \ref{fig:con_en:fig2}. 
Note that, all wireless stations are fixed, and both $S$ and $I$ are in transmit mode while AP being in receive mode. It is assumed that AP has a rotating drum antenna of radius, $R$, with the canister opening being directed towards the desired transmitter, $S$. Let the analog complex baseband signal of the desired and the interfering stations respectively be given by $b_S\left(t\right)$ and $b_I\left(t\right)$:
\begin{align}
b_S\left(t\right) &= \Psi\left(a_S\right),\\
b_I\left(t\right) &= \Psi\left(a_I\right),
\end{align} 
where $a_S$ and $a_I$ respectively are the complex discrete time base-band data--drawn from a $M$-Quadrature Amplitude Modulation ($M$-QAM) constellation, $\mathcal{M}$ based on binary data signals, $d_S$ and $d_I$--signal of stations, S and I and the operation $\Psi(.)$ denotes the squre root raised cosine (SRRC) pulse shaping operation \cite{Gold05}. Let the symbol time duration and the bandwidth of the data signals be denoted by $T_s$ and $B_w$ respectively, where typically $T_s=1/B_w$. The transmit waveforms will then be given by:
\begin{align}
x_S\left(t\right) &=\text{Re}\left\{b_S\left(t\right)e^{j2\pi f_ct}\right\},\\
x_I\left(t\right) &= \text{Re}\left\{b_I\left(t\right)e^{j2\pi f_ct}\right\},
\end{align}       
where $f_c$ is the carrier frequency, $j=\sqrt{-1}$, and $\text{Re}\left\{\right\}$ denotes the real part \cite{Stuber02}. Furthermore, the transmit power of both links are scaled to give $\mathcal{E}\left\{x_S (t)^2\right\}=P_S$ and $\mathcal{E}\left\{x_I (t)^2\right\}=P_I$. It is herein assumed that the communication takes place between $S$, $I$ and AP in a scattering environment, where AP receives multiple faded replicas of the transmitted signals, $x_S\left(t\right)$ and $x_I\left(t\right)$. Hence, the received signal by AP can in the absence of noise be given by:
\begin{align}
y\left(t\right) &= \text{Re}\left\{r_S\left(t\right)e^{j2\pi f_ct}\right\} +\text{Re}\left\{r_I\left(t\right)e^{j2\pi f_ct}\right\},
\end{align}
where the complex base-band desired and interfering received signals, $r_S\left(t\right)$ and $r_I\left(t\right)$ can be given by:
\begin{subequations}\label{com:receive:eq1}
\begin{align}
r_S\left(t\right)=\sum_{n=0}^{N_S}\alpha_{n}^S e^{j\phi_{n}^S(t)} b_S\left(t-\tau_{n}^S\right),\\
r_I\left(t\right)=\sum_{n=0}^{N_I}\alpha_{n}^I e^{j\phi_{n}^I(t)} b_I\left(t-\tau_{n}^I\right).
\end{align}
\end{subequations}
The quantities, $\alpha_{n}$, $\phi_{n}$ and $\tau_{n}$ in \eqref{com:receive:eq1} respectively are the faded amplitude, phase and path delay of the $n$th replica of the desired and interference signals, where:
\begin{subequations}\label{com:receive:eq2}
	\begin{align}
	\phi_{n}^S\left(t\right) &= 2\pi\left\{f_{n}^St- \left(f_c + f_{n}^S\right)\tau_{n}^S \right\},\\
	\phi_{n}^I\left(t\right) &= 2\pi\left\{f_{n}^It- \left(f_c + f_{n}^I\right)\tau_{n}^I \right\}.
	\end{align}
\end{subequations}
It is assumed that $\alpha_{n}^S$, $\alpha_{n}^I$, $\tau_{n}^S$, $\tau_{n}^I$, $N_S$ and $N_I$ are approximately the same for a certain amount of time (say block interval) that is sufficient to transmit at least one data packet, and change to new realizations independently in the next block interval. The frequency change due to Doppler effect on the $n$th incoming ray of the desired and the interfering signal, $f_{n}^S$ and $f_{n}^I$ in \eqref{com:receive:eq2} are respectively given by:
\begin{figure}[t]
	\centerline{\includegraphics*[scale=2.5]{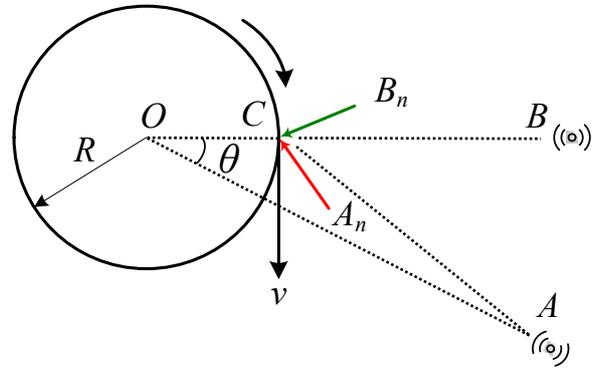}}
	\caption{The top view of an access point, where one desired (i.e. $B_nC$) and one interfering ray (i.e. $A_nC$) are shown. The canister opening is at C.}
	\label{fig:con_en:fig4}
\end{figure}  
\begin{subequations}\label{com:receive:eq3}
	\begin{align}
	f_{n}^S &= f_m \sin\left(\beta_{n}^S\right),\\
	f_{n}^I &= f_m \sin\left(\theta+\beta_{n}^I\right),
	\end{align}
\end{subequations}
where $f_m=f_cv/C$. Note that the arrival angles of users are measured with respect to the direction of the respective user. For instance, the arrival angles, $\beta_{n}^S$ of the desired user are measured with respect to the direction of CB (see Fig. \ref{fig:con_en:fig4}), and the arrival angles, $\beta_{n}^I$ of the interfering user are measured with respect to the direction of CA. As a result, according to Fig. \ref{fig:con_en:fig4}, the $n$th angle of arrival of the desired and the interfering signals are given by $\beta_{n}^S=\angle BCB_n$ and $\beta_{n}^I=\angle ACA_n$. Furthermore, the dominant angles of arrivals are limited to $-\omega/2 \leq \beta_n^S,\beta_n^I \leq \omega/2$ for all $n$. Since it is a narrow-band communication link, we further assume that $\tau_n^S,\tau_n^I \ll T_s$, and with out loss of applicability, make the substitution, $\tau_S \approx \tau_n^S$ and $\tau_I \approx \tau_n^I$ for all $n$. As a result:
\begin{subequations}\label{com:receive:eq4}
	\begin{align}
	r_S\left(t\right)=b_S\left(t-\tau_S\right)\sum_{n=0}^{N_S}\alpha_{n}^S e^{j\phi_{n}^S(t)},\\
	r_I\left(t\right)=b_I\left(t-\tau_I\right)\sum_{n=0}^{N_I}\alpha_{n}^I e^{j\phi_{n}^I(t)},
	\end{align}
\end{subequations}
One must note that the approximations, $\tau_S \approx \tau_n^S$ and $\tau_I \approx \tau_n^I$ are invoked only to $b_S\left(t-\tau_n^S\right)$ and $b_I\left(t-\tau_n^I\right)$ in \eqref{com:receive:eq4}, and since, $f_c\tau_n^S$ and $f_c\tau_n^I$ can still be significant, the path delay differences are still considered in the summation of \eqref{com:receive:eq4}. If the perfect synchronization is assumed, the complex baseband desired and interfering received signals, $r_S\left(t\right)$ and $r_I\left(t\right)$ can be rewritten as:
\begin{subequations}\label{com:receive:eq5}
	\begin{align}
	r_S\left(t\right)=\left\{\sum_{n=0}^{N_S}\alpha_{n}^S e^{j\phi_{n}^S(t)} \right\} b_S\left(t\right) = h_S\left(t\right) b_S\left(t\right), \\
	r_I\left(t\right)=\left\{\sum_{n=0}^{N_I}\alpha_{n}^I e^{j\phi_{n}^I(t)}\right\} b_I\left(t\right) =  h_I\left(t\right) b_I\left(t\right),
	\end{align}
\end{subequations}
where $h_S\left(t\right)$ and $h_I\left(t\right)$ are denoted henceforth as channel fading functions. The averaged channel gains for both links are defined as $\mathcal{E}\left\{|h_S (t)|^2\right\}=g_S$ and $\mathcal{E}\left\{|h_I (t)|^2\right\}=g_I$ \cite{Dush_general_jnl}. The constants, $g_S$ and $g_I$ capture the average channel gains due to path loss and shadowing alone which is also given by $g_S=\mathcal{E}\left\{\sum_{n=0}^{N_S} \left(\alpha_n^S\right)^2\right\}$ and $g_I=\mathcal{E}\left\{\sum_{n=0}^{N_I} \left(\alpha_n^I\right)^2\right\}$, where the expectation is over block intervals.\\ 
\indent It is assumed that dominant (in terms of the receive power) paths exist from both source and interferer to AP either as a result of line-of-sight (LoS) or dominant non-line-of-sight (NLoS) rays along with significantly weaker scattered rays. With out loss of generality, let the $0$th terms in \eqref{com:receive:eq1} denote the dominant paths, and as also pointed out earlier, AP points the canister opening towards dominant paths from $S$. Let $K$ be Rician $K$-factor which models the ratio of the received power between the dominant path and other paths \cite{Gold05}. Then:
\begin{align}
K &= \frac{\mathcal{E}\left\{\left(\alpha_0^S\right)^2\right\}}{\mathcal{E}\left\{\sum_{n=1}^{N_S} \left(\alpha_n^S\right)^2\right\}} = \frac{\mathcal{E}\left\{\left(\alpha_0^I\right)^2\right\}}{\mathcal{E}\left\{\sum_{n=0}^{N_I} \left(\alpha_n^I\right)^2\right\}},
\end{align}
where it is assumed that $K$-factor is the same for both the desired and interference link. The received signal in the presence of noise is given by:
\begin{align}
y\left(t\right) &= \text{Re}\left\{\left[r_S\left(t\right)+r_I\left(t\right)+n\left(t\right)\right]e^{j2\pi f_ct}\right\},
\end{align}
where $n\left(t\right)$ is complex base-band zero mean additive white Gaussian noise (AWGN) signal with $\mathcal{E}\left\{|n\left(t\right)|^2\right\}=\sigma_n^2$. The signal-to-noise-ratio is hence defined as $\text{SNR}=g_S P_S/\sigma_n^2$, and signal-to-interference power ratio is defined as $\text{SIR}=g_S P_S/g_I P_I$. The AP processes the received signal, $y\left(t\right)$ by in-phase and quadrature-phase mixing and filtering with a low pass filter (LPF) of bandwidth, $B_w$ to obtain the continuous-time complex base-band equivalent received signal as:
\begin{align}\label{com:receive:eq6}
r\left(t\right) &= r'_S\left(t\right)+r'_I\left(t\right)+n\left(t\right).
\end{align}
One must distinguish the difference between $r_S\left(t\right)$ and $r'_S\left(t\right)$ (also between $r_I\left(t\right)$ and $r'_I\left(t\right)$) in \eqref{com:receive:eq6} that $r'_S\left(t\right)$ is the low pass filtered version of $r_S\left(t\right)$ which is the original faded desired signal supposed to be received by AP. Conventionally, LPF assures that $r'_S\left(t\right)=r_S\left(t\right)$ and $r'_I\left(t\right)=r_I\left(t\right)$.
However as $v$ increases, and also discussed in detail in Sec. \ref{sec:the-effect-of-rotating-antenna}, $r_S\left(t\right)$ and $r_I\left(t\right)$ broaden in the frequency domain due to Doppler effect. Since the canister is directed towards the desired source, the spectral broadening in $r_S\left(t\right)$ is not severe, and under favorable fading conditions, reliable communication is still possible with reasonable channel estimation overhead. Moreover, if $v$ is sufficiently large, the spectrum of $r_I\left(t\right)$ shifts to an intermediate frequency determined by $v$ and $\theta$. Consequently, a majority or entire interference signal, $r_I\left(t\right)$, can be made to be filtered out by LPF so to create a less interfered channel. The AP samples $r\left(t\right)$ at symbol rate to obtain the discrete time complex base-band signal in terms of desired data signal, $a_S$ as:
\begin{subequations}\label{eq:received:A}
\begin{align}
r\left(\ell \right) &=  r'_S\left(\ell\right)+ n'\left(\ell\right), \\ &= h'_S\left(\ell\right) a_S\left(\ell \right)+ n'\left(\ell\right), \quad \forall \ell 
\end{align}
\end{subequations}
where $\ell$ alone is used for $\ell T_s$. Furthermore, $r'_S\left(\ell\right)$ and $n'\left(\ell\right)$ are the sampled versions of $r'_S\left(t\right)$ and $r'_I\left(t\right)+n\left(t\right)$ respectively. Note that $h'_S\left(\ell\right)$ combines the effect of $h_S\left(\ell\right)$ and other possible effects of low pass filtering of $r_S\left(\ell\right)$. The detector then uses the following symbol-by-symbol detection rule based on minimum Euclidean distance (also equivalent to maximum likelihood (ML) detector in AWGN) which treats the interference plus noise, $n'\left(\ell\right)$, as additional noise to obtain the estimated data, $\hat{d}_S$: 
\begin{align} \label{com:receive:eq7}
\hat{d}_S\left(\ell\right) &= \min_{a_S\left(\ell\right)\in \mathcal{M}} \left|r\left(\ell \right) - h'_S\left(\ell\right) a_S\left(\ell \right)\right|^2, \qquad \forall \ell.
\end{align}
Unlike in the case with $v=0$, due to Doppler effect, $h'_S\left(\ell\right)$ are different within a block interval even with $\alpha_{n}$, $\phi_{n}$, $\tau_{n}$, $N_S$ and $N_I$ being fixed. However, in this study, we assume that they can be approximated by a fixed value, $\hbar_S$. Consequently, \eqref{com:receive:eq7} becomes:
\begin{align} \label{com:receive:eq8}
\hat{d}_S\left(\ell\right) &= \min_{a_S\left(\ell\right)\in \mathcal{M}} \left|r\left(\ell \right) - \hat{\hbar}_S a_S\left(\ell \right)\right|^2, \qquad \forall \ell.
\end{align}
where $\hat{\hbar}_S$ is the estimated value of $\hbar_S$. The key roles played by spectral characteristics of channel fading functions in \eqref{com:receive:eq5} are graphically discussed in the next section.  
\subsection{The Effect of Rotating Antenna}\label{sec:the-effect-of-rotating-antenna}    
\begin{figure}[t]
\centerline{\includegraphics[scale=1.2]{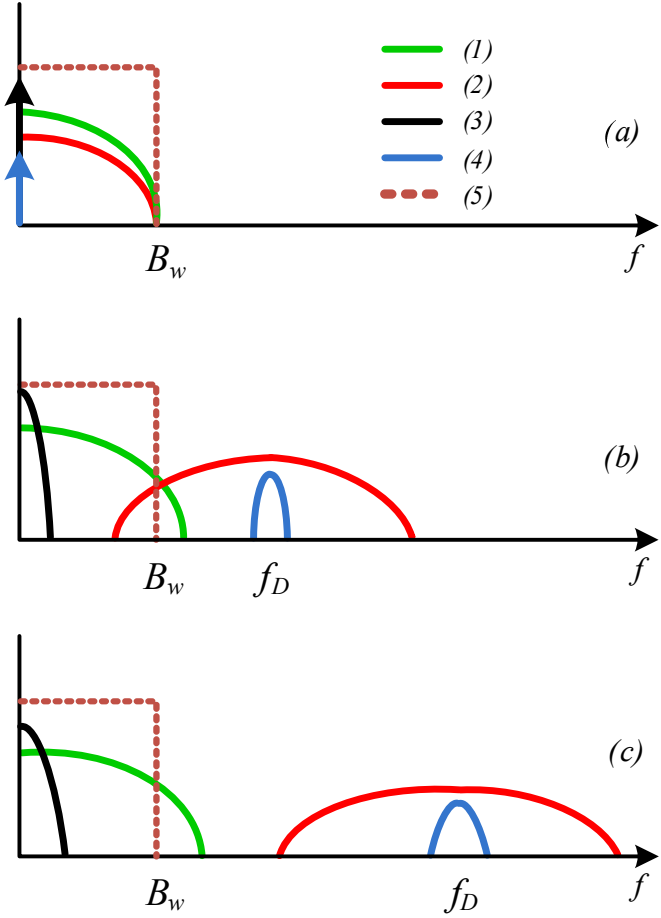}}
\caption{Illustrations of magnitude responses of (1) $R_S\left(f\right)$, (2) $R_I\left(f\right)$, (3) $H_S\left(f\right)$ and (4) $H_I\left(f\right)$ which are Fourier transforms of $r_S\left(t\right)$, $r_I\left(t\right)$, $h_S\left(t\right)$ and $h_I\left(t\right)$ respectively. The magnitude response of LPF at AP is also shown in (5), and $f_D=\frac{vf_c}{C}\sin\theta$.} 
\label{fig:con_en:fig5}
\end{figure}

Conventionally, the antenna is fixed (i.e. $v=0$), but as rotation speed increases, two conflicting phenomena happen. These phenomena can be better explained using the illustrations in Fig. \ref{fig:con_en:fig5}. The Fig. \ref{fig:con_en:fig5}-(a) shows an illustration of the single-sided magnitude response of $r_S\left(t\right)$, $r_I\left(t\right)$, $h_S\left(t\right)$ and $h_I\left(t\right)$ along with the magnitude response of the receiver's LPF. When $v=0$, $H_S\left(f\right)$ and $H_I\left(f\right)$ are just impulses, and have no relevant effect on $r_S\left(f\right)$ and $r_I\left(f\right)$. However, as $v$ increases $H_S\left(f\right)$ and $H_I\left(f\right)$ tend to broaden, and notably, $H_I\left(f\right)$ sways away from zero frequency (i.e.,$f=0$) to an intermediate frequency determined by $f_D=f_m\sin \theta$, and in turn by azimuth separation, $\theta$, $v$, and $f_c$. Consequently, the majority of interference power lies outside the desired signal bandwidth,  $B_w$, and hence, there is an interference suppression effect. On the other hand, since, $R_S\left(f\right)=H_S\left(f\right)\circledast B_S\left(f\right)$ and $R_I\left(f\right)=H_I\left(f\right)\circledast B_I\left(f\right)$, $R_S\left(f\right)$ and $R_I\left(f\right)$ also tend to broaden. Note that $B_S\left(f\right)$ and $B_I\left(f\right)$ denote the frequency response of $b_S\left(t\right)$ and $b_I\left(t\right)$ respectively, and $\circledast$ denotes the convolution operator \cite{Gold05}. As a result of this spectrum broadening\footnote{The spectral broadening is initiated by the rotation, but could be exacerbated by adverse fading conditions such as low $K$, and high $N_S$, $N_I$ and $\omega$.}, a certain amount of desired signal power is also suppressed by LPF and thus a distortion effect on the desired signal. As $v$ increases further, as shown in Fig. \ref{fig:con_en:fig5}-(c), the interference signal can be shifted completely away from the desired signal, but the amount of power suppressed by the LPF also increases making the desired signal more distorted. Hence, a trade off between interference suppression capability and the distortion of the desired signal in Doppler assisted wireless communication is clearly apparent. However, as shown in Sec. \ref{sec:simulation-results}, a reasonable compromise can be made, where significant performance gain can still be achieved.               
\section{Performance Analysis and Discussion}\label{sec:performance}
\begin{figure*}[t]
	\centering
	\includegraphics[scale=1.1]{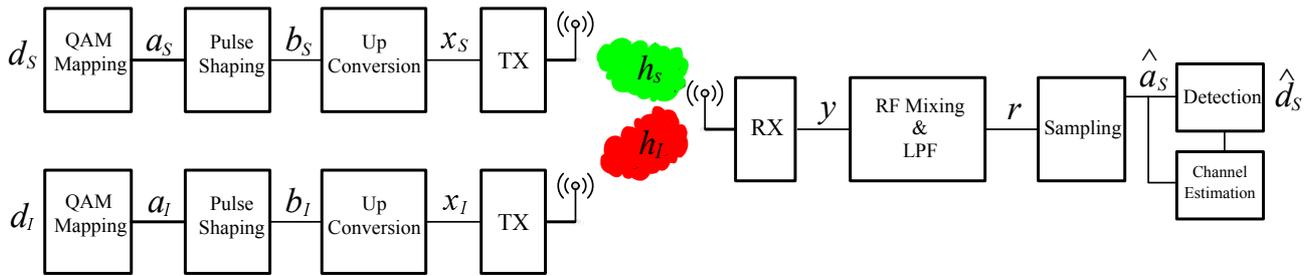}
	\caption{The main simulation block diagram.} 
	\label{fig:con_en:fig6}
\end{figure*}
The performance of DAIM is analyzed using a comprehensive end-to-end digital communication link simulated on MATLAB. In order to accurately assess DAIM, we herein simulate a pass-band digital communication link, where pulse shaping, up-conversion, RF mixing and LPF have also been implemented\footnote{Note that the implementation of up-conversion and RF mixing which requires a significantly higher sampling rate, and hence is computationally inefficient, is avoided by using an equivalent base-band model, but still with transmit pulse shaping and LPF in order to accurately captures the effects outlined in Sec. \ref{sec:the-effect-of-rotating-antenna}. Unlike in conventional complex base-band simulations, the LPF operation is crucial for this simulation study.}. The main block diagram of the simulation is shown in Fig. \ref{fig:con_en:fig6}, where for simplicity Quadrature Phase Shift Keying (QPSK) is considered with other system parameters as shown in Table. 1. The major steps of the simulation environment are obtained as follows. 
\subsection{Transmit Signals}
We consider a time duration to transmit a single data packet, where a single packet lasts $L$ symbols or equivalently $\eta L$ samples. The constant, $\eta$ denotes the up-sampling ratio, which is given by $\eta=T_s/t_o$, where $t_o=1/F_s$ is the sampling period in the computer simulation herein. The $k$th sample of the complex base-band transmitted signal of the desired link\footnote{Note that one can obtain the $k$th sample of the pass-band transmit signal of the desired link by $x_S\left(kt_o\right)=b_S\left(kt_o\right)\cos 2\pi f_ckt_o$.} is obtained by:
\begin{align}
b_S\left(kt_o\right) &=\left[\tilde{a}_S \circledast p\right]\left(kt_o\right), \qquad k=1,\dots,\eta L,
\end{align}
where $\tilde{a}_S$ is the $k$th sample of up-sampled version of $a_S$ and $p\left(kt_o\right)$ is the $k$th sample of SRRC filter which is obtained by:
\begin{align}
p\left(kt_o\right) &= \frac{\sin\left(\pi k\left(1-\rho\right)\right)+4\pi k\cos\left(\pi t \left(1+\rho\right)\right)}{\pi k \left(1-16k^2\rho^2\right)},
\end{align}
where $\rho$ is the roll-off factor of SRRC filter. Henceforth, we may interchangeably use standalone $k$ for $kt_o$. Furthermore, we scale $b_S\left(kt_o\right)$ so $\mathcal{E}\left\{|x_S\left(k\right)|^2\right\}=P_s/\eta=1/\eta$. Consequently, the $k$th sample of the normalized complex base-band faded desired received signal\footnote{Note that one can obtain the $k$th sample of the complex pass-band desired received signal by $\text{Re} \left\{r_S \left(k\right)e^{j2\pi f_ck}\right\}$.} is obtained by:
\begin{align}
\!\!\!\!r_S \left(k\right) &=b_S\left(k\right)h_S\left(k\right).
\end{align} 
\subsection{Multi-path Channels}\label{sec:multi-path-channels}
The $k$th sample of the fading function, $h_S\left(k\right)$ is obtained by the complex equation:
\begin{align}\label{sec3:eq1}
h_S\left(k\right) &= \sqrt{\frac{g_S K}{K+1}} \left[h_S\right]_d + \sqrt{\frac{g_S}{K+1}} \left[h_S\right]_s,
\end{align}
where the channel function of the direct path of the desired link, $\left[h_S\right]_d=e^{-j\varphi_0^S}$, and the channel function of the scattered paths, $\left[h_S\right]_s$ is obtained as: 
\begin{align}
\left[h_S\right]_s &= \sum_{n=1}^{N_S} \alpha_n^S e^{j2\pi f_n^Skt_o- j\psi_n^S}.
\end{align}
The term that accounts for the change in the frequency due to Doppler effect, $f_n ^S$ is obtained by $f_{n}^S =f_m \sin\left(\beta_{n}^S\right)$, where $\beta_n^S \sim \mathcal{U}\left(-\omega/2, \omega/2 \right)$. Note that $\mathcal{U}\left(a,b\right)$ is an abbreviation for the uniform distribution with support, $[a,b]$. The phase term, $\psi_n^S$ is obtained by $\psi_n^S \sim \mathcal{U}\left(0, 2\pi\right)$. More importantly, note that $f_0^S=0$ for any $v$ due to the fact that canister opening is directed towards to the desired transmitter. Furthermore, in LoS fading, $\psi_0^S$ is dependent on the distance between $S$ and AP, and hence is set to a fixed arbitrary value throughout the simulation. Lastly, the amplitudes, $\alpha_n^S$s are assumed to be approximately equal, and hence, are set to $\alpha_n^S=\sqrt{1/2 N_S}$ which in conjunction with \eqref{sec3:eq1} subsequently guarantees that $\mathcal{E}\left\{|h_S\left(k\right)|^2 \right\}=g_S$. This along with the fact that $\mathcal{E}\left\{|x_S\left(k\right)|^2 \right\}=1/\eta$ directly implies that $\mathcal{E}\left\{|r_S\left(k\right)|^2 \right\}=1/\eta$. Similarly, the $k$th sample of the scattered received signal of the interfering link, $r_I\left(kt_o\right)$ is also obtained with following notable exceptions: $\left[h_S\right]_d=e^{j2\pi f_0^I-j\psi_0^I}$, where $f_0^I=f_m \sin \theta$. Furthermore, $\beta_n^I$ and $\psi_n^I$ are assumed to be distributed as in the case for the desired link. 
\subsection{Receive Signal At AP}\label{sec:receive-signal}
As a result, the $k$th sample of the combined pass-band received signal by AP is obtained as:
\begin{align}
y\left(k\right) &= \text{Re} \left\{\left[r_S \left(k\right)+r_I \left(k\right) + n\left(k\right) \right] e^{j2\pi f_ck}\right\},
\end{align} 
where $n\left(k\right)$ is the $k$th AWGN sample with variance $\sigma_n^2/\eta$, and since $r_S \left(k\right)$ and $r_I\left(k\right)$ are normalized to have average channel gains, $g_S$ and $g_I$ respectively, SIR of the wireless network boils down to $\text{SIR}=g_S/g_I$, and can be adjusted conveniently by manipulating, $g_S$ and $g_I$ in the computer simulation herein. Furthermore, SNR also boils down to $\text{SNR}=g_S/\sigma_n^2$. 
\subsection{RF Mixing, LP Filtering, Sampling and Detection}\label{sec:rf-mixing,-lp-filtering,-sampling-and-detection}
It is assumed herein that AP performs I/Q mixing perfectly\footnote{Note that one can obtain the $k$th sample of the in-phase mixed signal by $y\left(k\right)\cos 2\pi f_ck$ while $y\left(k\right)\sin 2\pi f_ck$ being the quadrature phase mixed signal.}, and produces a base-band version of $y\left(k\right)$, which is $r_S \left(k\right)+r_I \left(k\right) + n\left(k\right)$. The AP then passes this complex base-band version of $y\left(k\right)$ through SRRC LPF. The low pass filtered complex signal is then sampled (rather down-sampled) at symbol rate of $T_s$ to obtain $r\left(\ell T_s\right),$ for $\ell=1,\dots, L$, which are the faded, interfered and noisier versions of the complex modulated samples, $a_S\left(\ell T_s\right),$ $\forall\text{ }\ell$. The $L$ complex samples per packet are then forwarded to the detector in \eqref{com:receive:eq8} to obtain the reproduced data, $\hat{d}_S$.      
\subsection{Simple Channel Estimation}\label{sec:simple-channel-estimation}
As pointed out in Sec. \ref{sec:system}, despite being different, all fading coefficients, $h'_S\left(\ell\right)$s, in a single data packet duration are approximated by a single value, $\hbar_S$. In this study, we assume that the desired transmitter sends $Q$ number of known data symbols, and AP uses simple least-square (LS) algorithm for channel estimation \cite{Haykin13}. From \eqref{eq:received:A}, the complex base-band signal received in the channel estimation phase, $r^e\left(\ell\right)$, is:
\begin{align}
r^e\left(\ell\right) &= h'_S\left(\ell\right) a^e_S\left(\ell \right) + n'\left(\ell\right), \quad \text{for} \quad \ell=1,\dots,Q,\\
&\approx \hbar_S a^e_S\left(\ell \right) + n'\left(\ell\right),
\end{align}
where $a^e_S\left(\ell \right)$ are known symbols transmitted for channel estimation. The $\text{LS}$ estimation of $\hbar_S$ can hence be obtained as: $\hat{\hbar}_S=\left(\bm{a}_S^e\right)^H\bm{r}^e/\left(\bm{a}_S^e\right)^H\bm{a}_S^e$, where $\bm{r}^e=\left\{r^e(1),\dots,r^e(Q)\right\}^T$ and $\bm{a}^e=\left\{a_S^e(1),\dots,a_S^e(Q)\right\}^T$. In the forthcoming simulation study, $Q=8$ is used.  
\subsection{Simulation Results}\label{sec:simulation-results}
\begin{figure}[t]
	\centerline{\includegraphics*[scale=0.65]{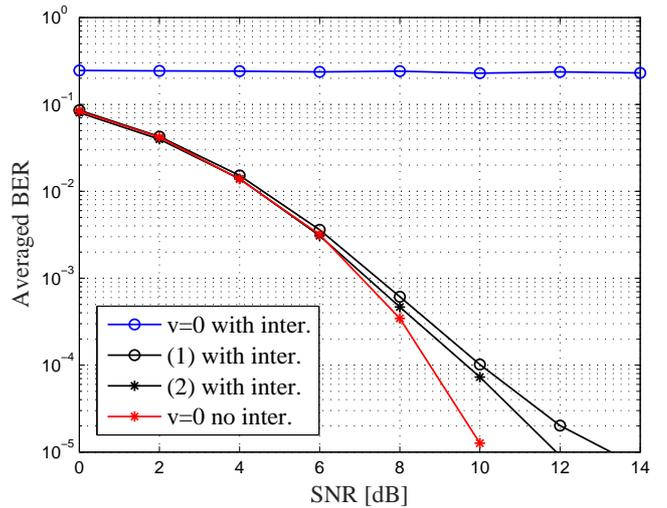}}
	\caption{BER performance of DAIM, where $\text{SIR}=0\text{ dB}$, $K=20\text{ dB}$, $N=50$, $\omega=20^o$, and $v$ is set such that in (1), $f_D=10.8\text{ KHz}$ and in (2), $f_D=12.9\text{ KHz}$.}
	\label{doppler:fig1}
\end{figure}

\begin{figure}[t]
	\centerline{\includegraphics*[scale=0.65]{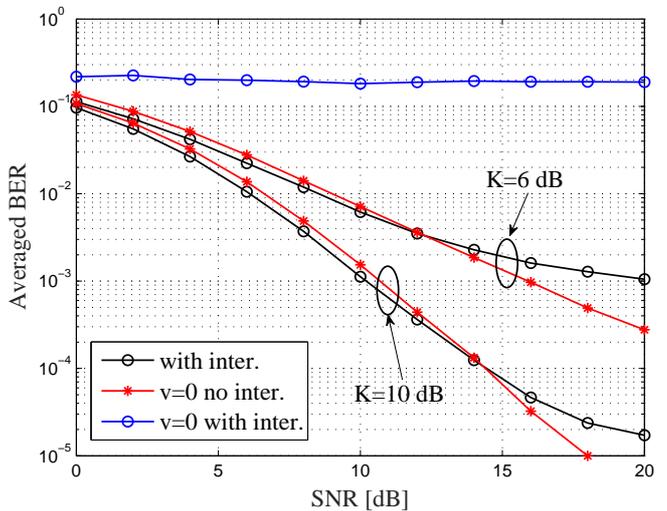}}
	\caption{BER performance of DAIM, where $\text{SIR}=0\text{ dB}$, $K=6/10\text{ dB}$, $N=20$, $\omega=10^o$ degrees, and $v$, when rotating, is set such that $f_D=17.3\text{KHz}$.}
	\label{doppler:fig2}
\end{figure}

\begin{figure}[t]
	\centerline{\includegraphics*[scale=0.65]{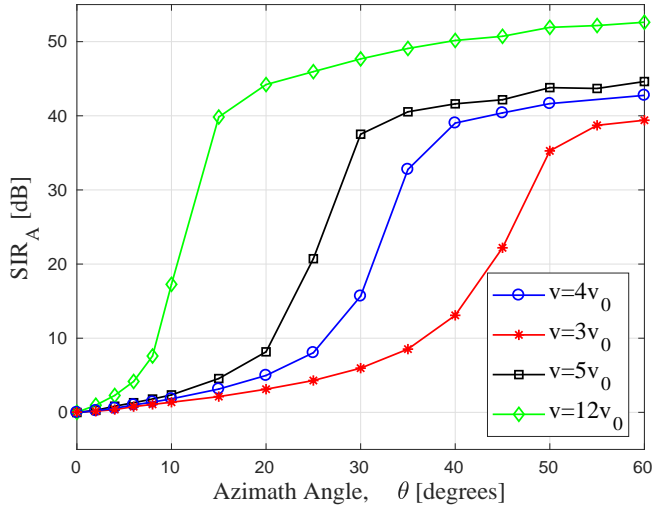}}
	\caption{$\text{SIR}_A$ performance of DAIM for $\text{SIR}=0\text{ dB}$, where $K=10\text{ dB}$, $N=20$, and $\omega=10^o$. Note that $v_0=\lambda_c B_w$.}
	\label{doppler:fig0}
\end{figure}

\begin{figure}[t]
	\centerline{\includegraphics*[scale=0.65]{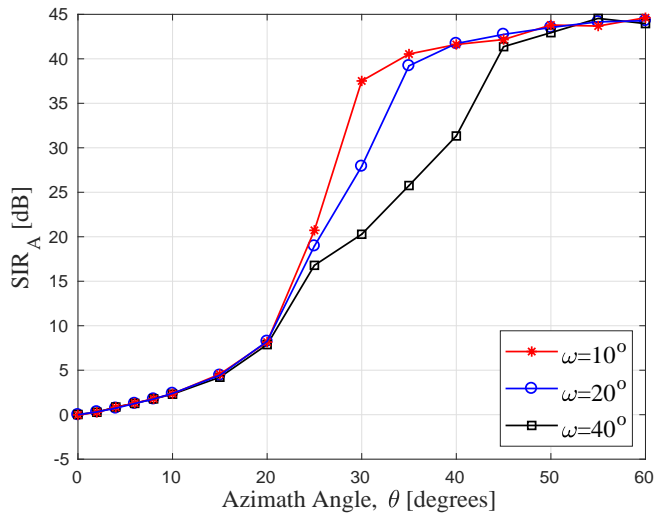}}
	\caption{$\text{SIR}_A$ performance of DAIM for $\text{SIR}=0\text{ dB}$, where $K=10\text{ dB}$, $N=20$, and $\omega=10^o/20^o/40^o$. Note that $v=5v_0$.}
	\label{doppler:fig01}
\end{figure}

\begin{figure}[t]
	\centerline{\includegraphics*[scale=0.65]{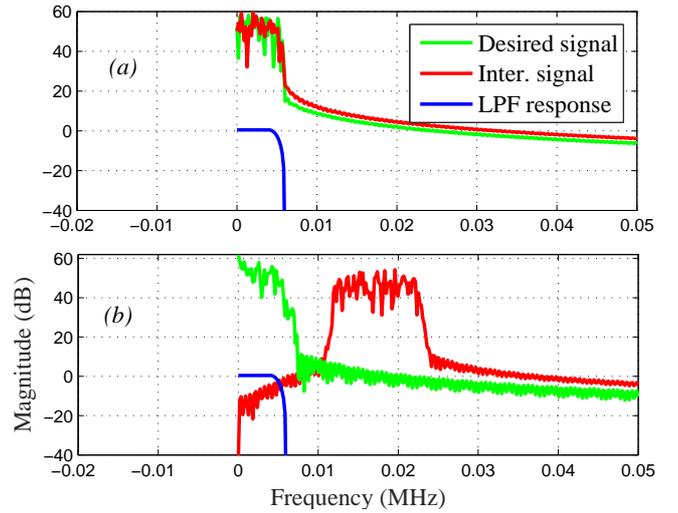}}
	\caption{Spectral characteristics of faded signals, $h_S\left(t\right)b_S\left(t\right)$ and $h_I\left(t\right)b_I\left(t\right)$, where $\text{SIR}=0\text{ dB}$, $K=10\text{ dB}$, $N=20$, $\omega=10^o$ degrees. In (a) $v=0$ and in (b), $v$ is set such that $f_D=17.3\text{ KHz}$.}
	\label{doppler:fig3}
\end{figure}

\begin{table}[t]
	\caption{Parameters For QPSK Pass-band Simulation}
	\centering
	\begin{tabular}{| l | c |}
		\hline \hline
		Parameter & Value \\
		\hline
		Carrier frequency, $f_c$ & 60 GHz\\
		Sampling frequency, $F_s$ & 3 MHz \\
		Signal bandwidth, $B_w$ & 5 KHz \\
		Symbol duration, $T_s$ & $1/B_w$\\
		Low pass filter & SRRC \\
		SRRC span & $64T_s$ \\
		Rolloff factor of SRRC, $\rho$ & 0.2 \\
		Over sampling rate, $\eta$, & $300$\\
		Packet Length, $L$ & $500T_s$\\
		\hline
	\end{tabular}\label{table:sm:1}
\end{table}
A severely interfered link is simulated, where both desired and interference links have equal average link gains, so $g_I=g_S$. Hence, the SIR before the DAIM receiver denoted herein as $\text{SIR}$ is 0 dB. Fig. \ref{doppler:fig1} shows the averaged BER performance of a communication link with $\text{SIR}=0 \text{ dB}$, $N_S=N_I=50$, $K=20\text{ dB}$, and $\omega=20^o$, where the results show that when $v=0$, the link with interference is completely unusable. However, as $v$ increases to $v=2.5\lambda_cB_w$, BER performance improves significantly. BER performance with no interference and $v=0$ is also shown for comparison. It is apparent that at low SNR, Doppler assisted communication can create an interference free link, but as SNR increases, BER performance drifts away. This trend can be attributed to the phenomenon that as $v$ increases, the spectrum of the desired signal $h_S\left(t\right)b_S\left(t\right)$ broadens, which in turn makes a certain amount of desired signal power suppressed by LPF at AP.\\
\indent  Fig. \ref{doppler:fig2} shows BER performance of the proposed interference mitigation system with $\text{SIR}=0\text{ dB}$, $N=20$, $K=6/10 \text{ dB}$ and $\omega=10^o$. As $v$ increases, similar to Fig. \ref{doppler:fig1}, BER performance significantly improves specially at low SNR. As SNR increases BER performance again drifts away from BER performance of the completely interference free link. In this fading condition, two major factors come into effect. The first one is the effect that in low $K$ values, the spectral broadening of $h_S\left(t\right)$ is severe, and hence a relatively larger amount of power is suppressed by LPF. The second one is the channel estimation errors. In the absence of significantly dominant multi-path components, the volatility of $h_S\left(t\right)$ even in the time duration of a single packet may be considerable. Approximating $h_S\left(\ell\right)$ for all $\ell$s by a single $\hbar_S$ is obviously suboptimal, and hence, more tailored algorithms for desired channel estimation may be needed. Furthermore, it is anticipated that more scenario specific low pass filters that passes a majority of desired signal power will be more effective for the earlier challenge as well.\\
\indent Fig. \ref{doppler:fig2} also shows that BER of DAIM is marginally better than BER of interference free link with $v=0$. This gain, though small, is defined as \textit{Doppler Gain (DG)}. Doppler effect causes the fading function, $h_S\left(t\right)$, to fluctuate specially in low $K$ and high $N_S$ and $\omega$. As a result, certain fading coefficients, $h_S\left(\ell\right)$, enhance their respective data symbols. Consequently, a net BER gain, which is manifested in the BER performance as DG, can be achieved. The simulation results that do not appear in this paper for reasons of space also show that ideal channel state information of the desired link significantly increases both BER of DAIM and DG.\\
\indent Fig. \ref{doppler:fig0} shows another view point of CCI mitigation capability of Doppler assisted wireless communication. Let the SIR after DAIM be denoted by $\text{SIR}_A$, and it is given by:
\begin{align}
\text{SIR}_A &= \frac{\mathcal{E}\left\{|r'_S(t)|^2\right\}}{\mathcal{E}\left\{|r'_I (t)|^2\right\}},
\end{align}
where $r'_S(t)$ and $r'_I(t)$ are given in \eqref{com:receive:eq6}. Fig. \ref{doppler:fig0} shows the $\text{SIR}_A$ performance of DAIM against the azimuthal separation, $\theta$ for different rotation speeds. It can be seen here that higher rotation speed may be required to achieve a certain $\text{SIR}_A$ for low $\theta$s and vice versa. Furthermore, Fig. \ref{doppler:fig01} shows $\text{SIR}_A$ performance of DAIM for various angle spreads ($\omega=10^o/20^o/40^o$) for a fixed rotation speed of $v=5v_0$, where $v_0=\lambda_c B_w$.\\ 
\indent Fig. \ref{doppler:fig3}-(a) and (b) show the spectral characteristics of faded signals, $h_S\left(t\right)b_S\left(t\right)$ and $h_I\left(t\right)b_I\left(t\right)$ with static and rotating antenna respectively. Herein, we consider a fading scenario, where $\text{SIR}=0\text{ dB}$, $K=10\text{ dB}$, $N=20$, $\omega=10^o$ degrees, and $v$ is set, when rotating, such that $f_D=17.3\text{ KHz}$. The figure clearly shows that as $v$ increases, the interference signal, $h_I\left(t\right)b_I\left(t\right)$ shifts to an intermediate frequency so that it is suppressed by LPF. Furthermore, the spectral broadening, as $v$ increases, of the desired signal is also visible in Fig. \ref{doppler:fig3}.            
\subsection{Important Remarks}\label{sec:remarks}
As shown in Sec. \ref{sec:simulation-results}, DAIM suppresses co-channel interference significantly as $v$ increases. The optimum $v$ is dependent on many system parameters such as $B_w$, $f_c$ and environmental and topological parameters such as $\theta$, $\omega$, and $K$. Hence, $v$ should be carefully selected, and be able to be adapted to the environment.  From Fig. \ref{doppler:fig1} and also in general, a rotation velocity that achieves $f_D=2B_w$ (which is about $v=2\lambda_cB_w\sin \theta$) is a reasonable value for $v$. It is equivalent to $v=58 \text{ m/s}$ for $B_w=5\text{ KHz}$, and $v=23 \text{ m/s}$ for $B_w=2\text{ KHz}$. From geometry of the drum antenna, the angular rotation speed can be obtained as: 
\begin{align}\label{rpm:eq1}
s_r &= \frac{30v}{\pi R} = \frac{60 C B_w}{\pi f_c R}\qquad \text{RPM},
\end{align}
where $s_r$ is the angular rotation speed in rounds per minute (RPM), which is about $1100\text{ RPM}$ for $B_w=2\text{ KHz}$, $R=20\text{ cm}$ and $f_c=60\text{ GHz}$. The equation, \eqref{rpm:eq1} also shows that $R$ and $s_r$ can be traded-off for one another. Furthermore, it appears that DAIM can only be applied practically for mmWave frequencies with $B_w$ in the order of $\text{KHz}$ and ultra-narrowband (UNB) communication systems with sub-GHz carrier frequencies with $B_w$ in the order of $\text{Hz}$. Other scenarios may require extremely high rotation velocities which may not be practically realizable with today's technologies. Otherwise, the results presented in this paper theoretically hold for any system that satisfies the assumptions considered in this paper.\\
\indent In wireless communication, all the multi-path signals contribute to the receive signal power. As rotation speed increases some desired multi-path signals that give rise to excessive Doppler shift could also (while, of course, suppressing majority of interfering multi-path signals) be suppressed out by the low pass filter (See Fig. \ref{fig:con_en:fig5}-(b) and (c)). It is important to note herein that, in the proper and advanced design of Doppler effect assisted wireless communication systems, the choice of the rotating speed should strike an effective balance between suppressing the interfering multi-paths and the desired multi-paths.    
\section{Further Remarks}\label{sec:further-remarks}
\subsection{Multi-antenna Configurations}
\begin{figure}[t]
	\centerline{\includegraphics*[scale=1.9]{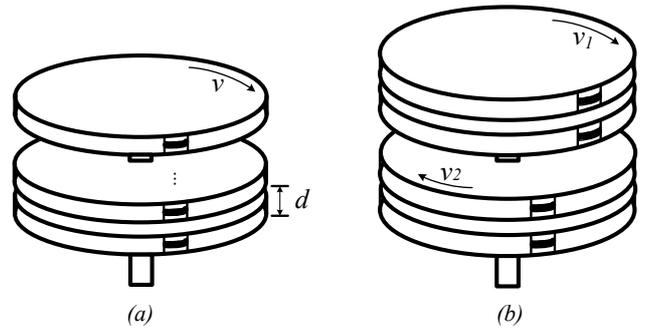}}
	\caption{(a) multi-antenna system for single-user communication and (b) multi-antenna system for multi-user configuration, where a setting for two-user system is shown to reduce the clutter.}
	\label{doppler:fig4}
\end{figure}

\begin{figure}[t]
	\centerline{\includegraphics*[scale=1.9]{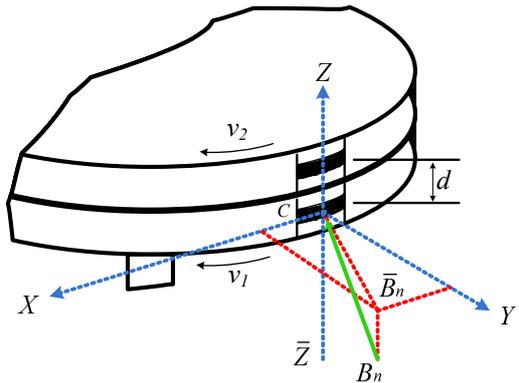}}
	\caption{An enlarged multi-antenna part-canister system that shows the azimuth and elevation of incoming rays. Only a single desired incoming ray, $B_n C$, is shown to reduce the clutter, where $B_nC\bar{Z}=\phi_n^S$ and $\bar{B}_nCY=\beta_n^S$}
	\label{doppler:fig45}
\end{figure}

\begin{figure}[t]
	\centerline{\includegraphics*[scale=0.65]{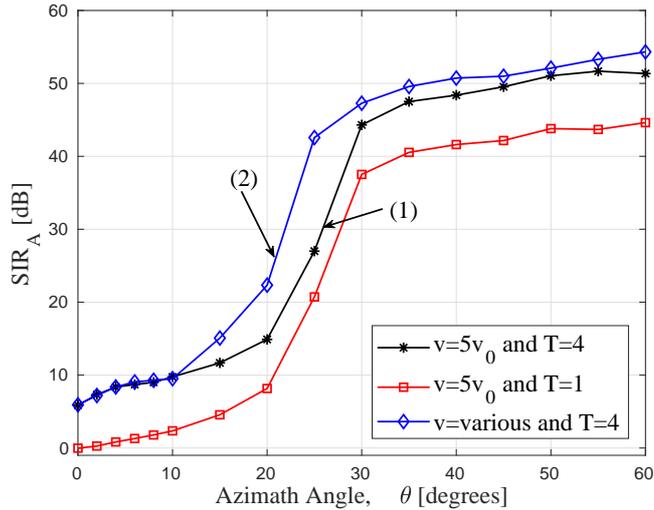}}
	\caption{$\text{SIR}_A$ performance of multi-antenna DAIM for $\text{SIR}=0\text{ dB}$, where $K=10\text{ dB}$, $N=20$, and $\omega=10^o$. Note that $v_0=\lambda_c B_w$.}
	\label{doppler:fig6}
\end{figure}

Doppler assisted communication systems can also be extended to accommodate multiple antennas and users as shown in Fig. \ref{doppler:fig4}, where a possible configuration for multi-antenna Doppler assisted system for single-user communication is shown in Fig. \ref{doppler:fig4}-(a). Extending \eqref{com:receive:eq6} to a dual-antenna configuration (merely for simplicity, but readily extends to more than two antenna cases) give rise to following base-band analogue equations:
\begin{subequations}\label{com:receive:eq9}
\begin{align}
r_1\left(t\right) &= r'_{S1}\left(t\right)+n'_1\left(t\right), \\
r_2\left(t\right) &= r'_{S2}\left(t\right)+n'_2\left(t\right).
\end{align}
\end{subequations}
Note that \eqref{com:receive:eq9} applies after low-pass filtering, and hence $r'_{Si}\left(t\right)=h'_{Si}\left(t\right)b_{S}\left(t\right)$ for $i=1,2$. Note herein that $h'_{Si}\left(t\right) \neq h_{Si}\left(t\right)$ due to Doppler effect and subsequent low-pass filtering. As shown in Fig. \ref{doppler:fig45}, the canisters are stacked vertically, and hence both elevation and the azimuth angle of the incoming rays are considered in this simulation. Consequently\footnote{It is assumed herein that unit wave vector of the $n$th desired wave front is given by $\sin \phi_n^S\sin \beta_n^S \underbar{i}+\sin \phi_n^S\cos \beta_n^S \underbar{j}+\cos\phi_n^S \underbar{k}$, and antenna velocity vector of the $i$th drum antenna is $v_i\underbar{i}$. See fig. \ref{doppler:fig45} for an illustration.},
\begin{align}
h_{Si}(t) &= \sum_{n=0}^{N_S}\alpha_n^S e^{j2\pi f_n^St-j\psi_n^S+j(i-1)d\cos\phi_n^S},
\end{align}
where $f_n^S=v_i\frac{f_c}{C}\sin\phi_n^S\sin\beta_n^S$, and $\psi_n^S \sim \mathcal{U}\left(0,2\pi\right)$.  Note that $\beta_n^S$ and $\phi_n^S$ respectively are azimuth and elevation of the $n$th incoming ray measured in anti-clockwise direction with respect to $CY$ and $C\bar{Z}$ axis respectively (see Fig. \ref{doppler:fig45}). Also, $\beta_n^S \sim \mathcal{U}\left(-\omega/2, \omega/2 \right)$, and $\phi_n^S \sim \mathcal{U}\left(\pi/2-\omega/2, \pi/2+\omega/2 \right)$, where it is assumed that both azimuth and elevation spread are the same. Note that $0$th path denotes the dominant multi-path, and hence, $\beta_0^S=0$, and it is also assumed that $\phi_0^S=1.39626$ which is $80^0$degrees. Similar fashion, $h_{Ii}(t)$ can also be obtained with the notable exception of $f_n^I=v_i\frac{f_c}{C}\sin\phi_n^I\sin(\theta+\beta_n^I)$. The ideal maximum ratio combining (MRC) can be achieved by:
\begin{align}
\hat{r}\left(t\right) &= \left(\bm{h}'_{S}\left(t\right)\right)^H \bm{r}\left(t\right),
\end{align}
where $\hat{r}(t)$ is the combiner output and $\left(\bm{h}'_{S}\left(t\right)\right)^H$ is the Hermitian conjugate of $\bm{h}'_{S}\left(t\right)$, $\bm{h}'_{S}\left(t\right)=\left\{ h'_{S1}\left(t\right)  h'_{S2}\left(t\right) \right\}^T$ and also $\bm{r}\left(t\right)=\left\{r_{1}\left(t\right) r_{2}\left(t\right)\right\}^T$. However, often $h'_{Si}\left(t\right)$ cannot be estimated exactly, and one reasonable remedy is to use $\hat{\hbar}_{Si}$ which is also used for data detection in \eqref{com:receive:eq8}, and note that the multi-antenna DAIM simulations in this section also employ $\hat{\hbar}_{Si}$. Fig. \ref{doppler:fig6} shows $\text{SIR}_A$ performance after DAIM processing and combining, where $T$ is the number of drum antennas configured as shown in Fig. \ref{doppler:fig4}-(a). It is clear that $\text{SIR}_A$ improves significantly as $T$ increases, and interestingly, one can manipulate the number of antennas, $T$, and the rotation speed, $v$, in order to achieve a certain $\text{SIR}_A$ performance. It is anticipated that better channel estimation techniques shall increase $\text{SIR}_A$ further. Note that all the drum antennas rotate at the same speed ($v=5v_0$) for curve (1) in Fig. \ref{doppler:fig6} which is not a necessary requirement.\\
\indent The curve (2) in Fig. \ref{doppler:fig6} shows the $\text{SIR}_A$ performance of a multi-antenna DAIM system with drums being rotated at different speeds of $v=3v_0, 4v_0, 5v_0, \text{and } 6v_0$. Let this speed profile be denoted as $\mathcal{SP}_2=\left\{3v_0, 4v_0, 5v_0, 6v_0\right\}$. It is clear that $\text{SIR}_A$ performance is always better than or the same as the case with drums being rotated at the same speed (i.e. a speed profile of $\mathcal{SP}_1=\left\{v=5v_0, 5v_0, 5v_0, 5v_0\right\}$). The energy required to rotate a single drum is proportional to the square of its angular velocity\footnote{This is due to the fact that kinetic energy required to rotate a rigid body at certain angular velocity is $0.5 I\omega^2$, where $I$ is the moment of inertia and the angular velocity respectively.} and in turn to the square of $v$. Hence, total energy, $E_T$ required to rotate drum antennas in different profiles are: 
\begin{align}
E_T \propto \begin{cases}
100 v_0^2 & \mathcal{SP}_1,\\
86 v_0^2 & \mathcal{SP}_2,
\end{cases}
\end{align}
and from kinetic energy efficiency perspective, $\mathcal{SP}_2$ is more preferable as it is $14\%$ more energy efficient in comparison with  $\mathcal{SP}_1$.    
\subsection{Future Research}
Even though the canister opening is directed to the direction of dominant scatters from the desired user, some non-dominant scatters can still induce fluctuations in the desired channel. Though these fluctuations can be harnessed to obtain some diversity gain as discussed in Sec. \ref{sec:simulation-results}, some deployments might prefer minimal channel fluctuations. Very closely vertically placed oppositely rotating drum antennas could be used to reduce Doppler induced channel fluctuations. Note that this approach may not reduce Doppler shift in all fading conditions, and more research is required to understand the depth and breadth of this approach. Furthermore, as shown in Fig. \ref{doppler:fig4}-(b), multi-antenna configuration can also be used for multi-user communication.\\
\indent The current paper discusses the basic operation of Doppler Assisted Wireless Communication, and demonstrates the feasibility of it for CCI mitigation. We have herein used standard modulation techniques, channel estimation techniques, pulse shaping and filtering methods just to demonstrate the feasibility of the proposed system. It is expected that more tailored data modulation techniques \cite{Dean17}, pulse shaping/filtering and also channel estimation techniques \cite{Stoica05,ZhGe17} will increase the robustness and the performance of the proposed scheme.\\
\indent It is also important to study other use cases of DAIM. In this paper, we have assumed interference occurs from a single interferer, but if interference occurs from unknown number of interferers from unknown locations spread over a large azimuth, the state-of-the-art techniques like MIMO can be very ineffective due to their high reliance on CSI. However, DAIM, in these type of extremely hostile environments could be very effective.    
\section{Conclusions}\label{sec:conclusions}
The current paper have introduced, and studied a new class of systems termed as Doppler assisted wireless communication. The proposed class of systems employs rotating drum antennas, and exploits Doppler effect, kinetic energy, and the topological information of wireless networks for co-channel interference mitigation. This paper includes a detailed simulation study that models several important system and environmental parameters. The results presented herein show that difficult co-channel interference--in the sense that it is statistically no more or less stronger to the desired signal, and often poorly handled by existing interference mitigation techniques--can successfully be mitigated by the proposed system. This paper has also discussed several important phenomena occurred in Doppler assisted communication systems such as Doppler gain along with advantages and challenges of Doppler assisted interference mitigation.      
\balance

\end{document}